\documentclass[useAMS,usenatbib]{mn2e}
\usepackage{graphicx}

\def\primip{\hbox{\rlap{\hbox{.}}\hbox{$'$}}}
\def\gradip{\hbox{\rlap{\hbox{.}}\raise 5.truept \hbox{{\small $\circ$}}}}

\title[The old metal-poor open cluster ESO 92-SC05]
{The old metal-poor open
 cluster ESO 92-SC05: accreted from a dwarf
galaxy?}
\author[S. Ortolani, E. Bica, and B. Barbuy]{S. Ortolani$^{1}$\thanks{E-mail:
sergio.ortolani@unipd.it; Observations collected at the
European Southern Observatory (ESO), La Silla, Chile,
proposal n$^{\circ}$ 64.L-0212(A)}; 
E. Bica$^{2}$\thanks{E-mail: bica@if.ufrgs.br};
B. Barbuy$^{3}$\thanks{E-mail: barbuy@astro.iag.usp.br}\\
$^{1}$Universit\`a di Padova, Dipartimento di Astronomia, 
Vicolo dell'Osservatorio 2, I-35122 Padova, Italy\\
$^{2}$Universidade Federal do Rio Grande do Sul, Dept. de Astronomia,
CP 15051, Porto Alegre 91500-970, Brazil\\
$^{3}$Universidade de S\~ao Paulo, Rua do Mat\~ao 1226, 05508-900
 S\~ao Paulo, Brazil}
\begin{document}

\date{}

\pagerange{\pageref{firstpage}--\pageref{lastpage}} \pubyear{2002}

\maketitle

\label{firstpage}

\begin{abstract}
The study of old open clusters outside the solar circle can bring
constraints on formation scenarios of the outer disk. In particular,
accretion of dwarf galaxies has been proposed as a likely mechanism
in the area.  
We use BVI photometry for determining fundamental parameters of the faint
open cluster ESO 92-SC05.
Colour-Magnitude Diagrams are compared with Padova isochrones,
in order to derive age, reddening and distance.
We derive a reddening E(B-V)= 0.17, and an old age of $\sim$6.0 Gyr.
 It is one of the rare open clusters
known to be older than 5 Gyr. A metallicity of Z$\sim$0.004 or 
[M/H]$\sim$-0.7 is found.  
The rather low metallicity suggests that this cluster might be
the result of an accretion episode of a dwarf galaxy.
\end{abstract}

\begin{keywords}
open clusters and associations: individual: 
ESO 92-SC05 - HR diagram
\end{keywords}

\section{Introduction}

The population of old Galactic
open clusters was reviewed by Friel (1995), where it was
pointed out the importance of studies
of individual clusters to better understand stellar
and dynamical evolution. New objects  help constrain
disc abundance gradients,  the chemical enrichment
and mixing in the disc, and age-metallicity relation.

Dias et al. (2002)\footnote{http://www.astro.iag.usp.br/$\sim$wilton/} 
reported 1537 open clusters of all 
ages as well as  additional open cluster candidates. 
 Both the Dias et al. study and the WEBDA database 
(Mermilliod 1996)\footnote{http://www.univie.ac.at/webda/}
 have compiled the  basic parameters of known
clusters  from the literature. 
Janes \& Phelps (1994) and Friel (1995) have defined  old
open clusters as those with  ages older than the Hyades ($\sim$700 Myr).
Such objects are sometimes referred
to as intermediate age clusters (IAC).
 A total of  108 confirmed old open clusters 
 was reported in Ortolani et al. (2005b), where 
 the age distribution  of open clusters was analysed.

Janes \& Phelps (1994) analysed a sample of 72 old open clusters
showing evidence that the Galactic disk might have been constantly
disturbed by infalling material. More recently, Frinchaboy et al. (2004)
suggested that part of the outer old open clusters may have formed in 
an accreted dwarf galaxy. Finally, Rocha-Pinto et al. (2006) find evidence
of a disrupted accreted dwarf galaxy  called Argo.
Bonatto et al. (2006) have shown that open clusters older than 1 Gyr
in their sample reach
heights up to Z$_{\rm GC}$=350 pc. Disc heating could 
be efficient for thickening
the disc in terms of field stars, but it is not established
 how old open clusters
might acquire vertical velocity for such heights. 

In this work we study the open cluster ESO 92-SC05 located in the fourth
quadrant, that presents an unusually low metallicity and an old age. 
The cluster ESO 92-SC05 was discovered in the ESO blue plates
(Lauberts 1982 and references therein).
It has coordinates J2000.0 $\alpha$ = 10$^{\rm h}$03$^{\rm m}$14${\rm s}$ 
and $\delta$ = -64$^{\rm o}$45'12" (l = 286$\gradip$19,
b =  -7$\gradip$50).
In the cluster direction  Schlegel et al. (1998) give a reddening
E(B-V)=0.20.

In Sect. 2 the observations are described. In Sect. 3 
the Colour-Magnitude Diagrams (CMD) are presented.
In Sect. 4 we derive cluster parameters and discuss possibilities
for the origin of such an old and metal-poor star cluster.
Concluding remarks are given in Sect. 5.

\begin{figure}
\resizebox{\hsize}{!}{\includegraphics{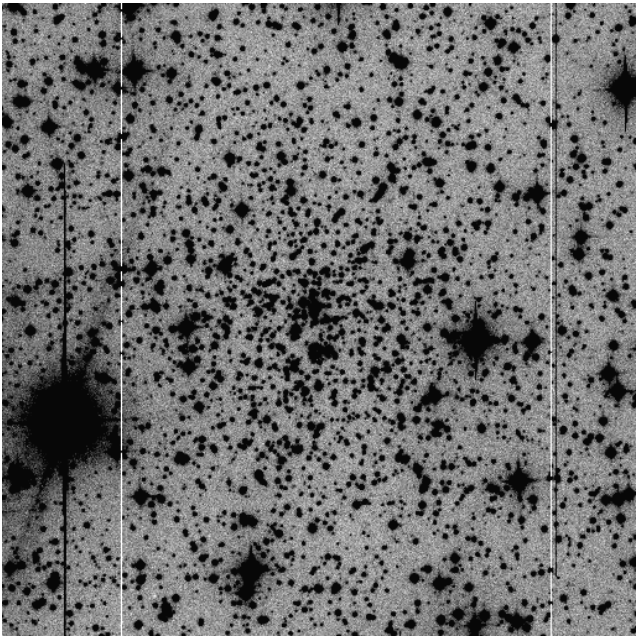}}
  \caption{
$V$ image (15 min) of ESO 92-SC05.
The field extraction size is 
6$\primip$3$\times$6$\primip$3. 
North is up and east to the left.
}
\label{fig1}
\end{figure}

\section { Observations and reductions} 

ESO~92-SC05 was observed on 2000 March 5
with the 1.54m Danish telescope  at ESO (La Silla). 
A Loral/Lesser CCD detector C1W7 with 2052$\times$2052 pixels, 
of pixel size 15 $\mu$m was used. It corresponds 
to $0.39"$ on the sky, providing a full 
field of $13'\times 13'$. 
The log of observations is reported in Table 1.
In Fig. \ref{fig1} is shown a 15 min  $V$ exposure of ESO~92-SC05
for a field extraction of  6.5'$\times$6.5' 
(1000$\times$1000 pixels).  
The image suggests  a core 
concentration of stars surrounded by a halo of fainter stars.
DAOPHOT II (Stetson 1992 and references therein)
 was used to extract the instrumental magnitudes. 
For calibrations we used  Landolt (1983, 1992) standard stars.

The calibration equations,  are:

$V$ = 26.46 + 0.01 ($B-V$) + $v$

$B$ = 26.40 + 0.1 ($B-V$) + $b$

\begin{figure}
\resizebox{\hsize}{!}{\includegraphics[angle=-90]{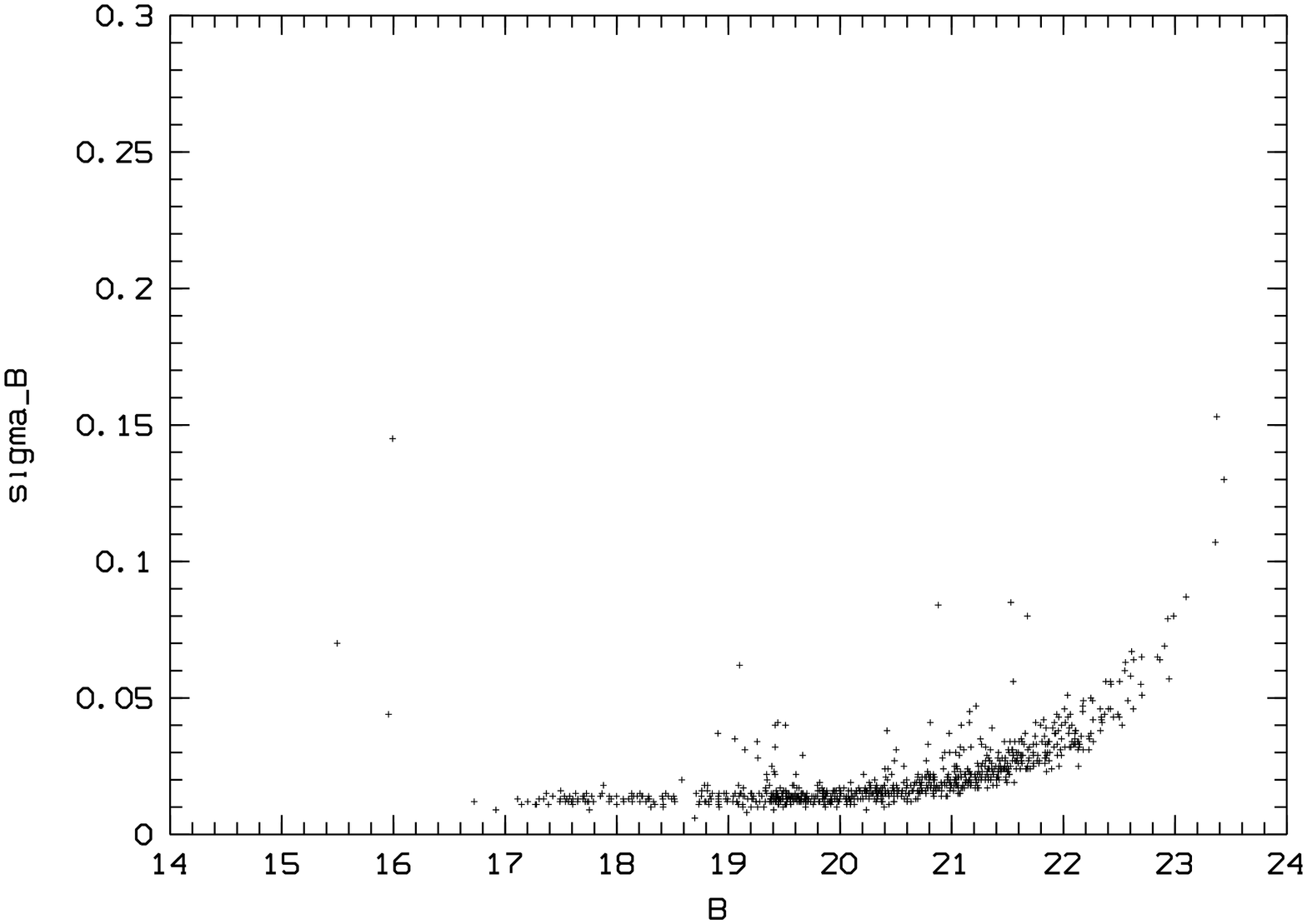}}
\resizebox{\hsize}{!}{\includegraphics[angle=-90]{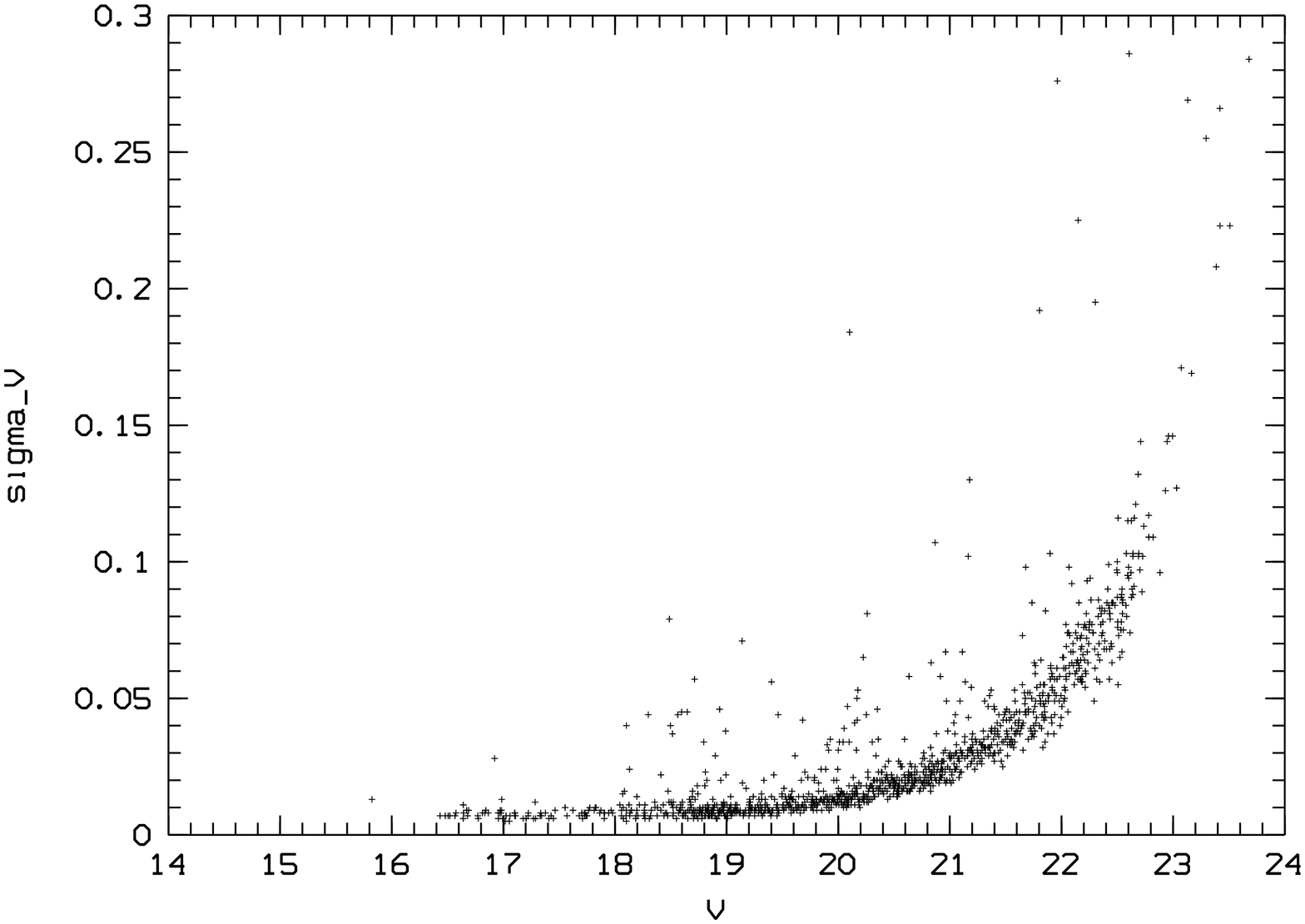}}
\resizebox{\hsize}{!}{\includegraphics[angle=-90]{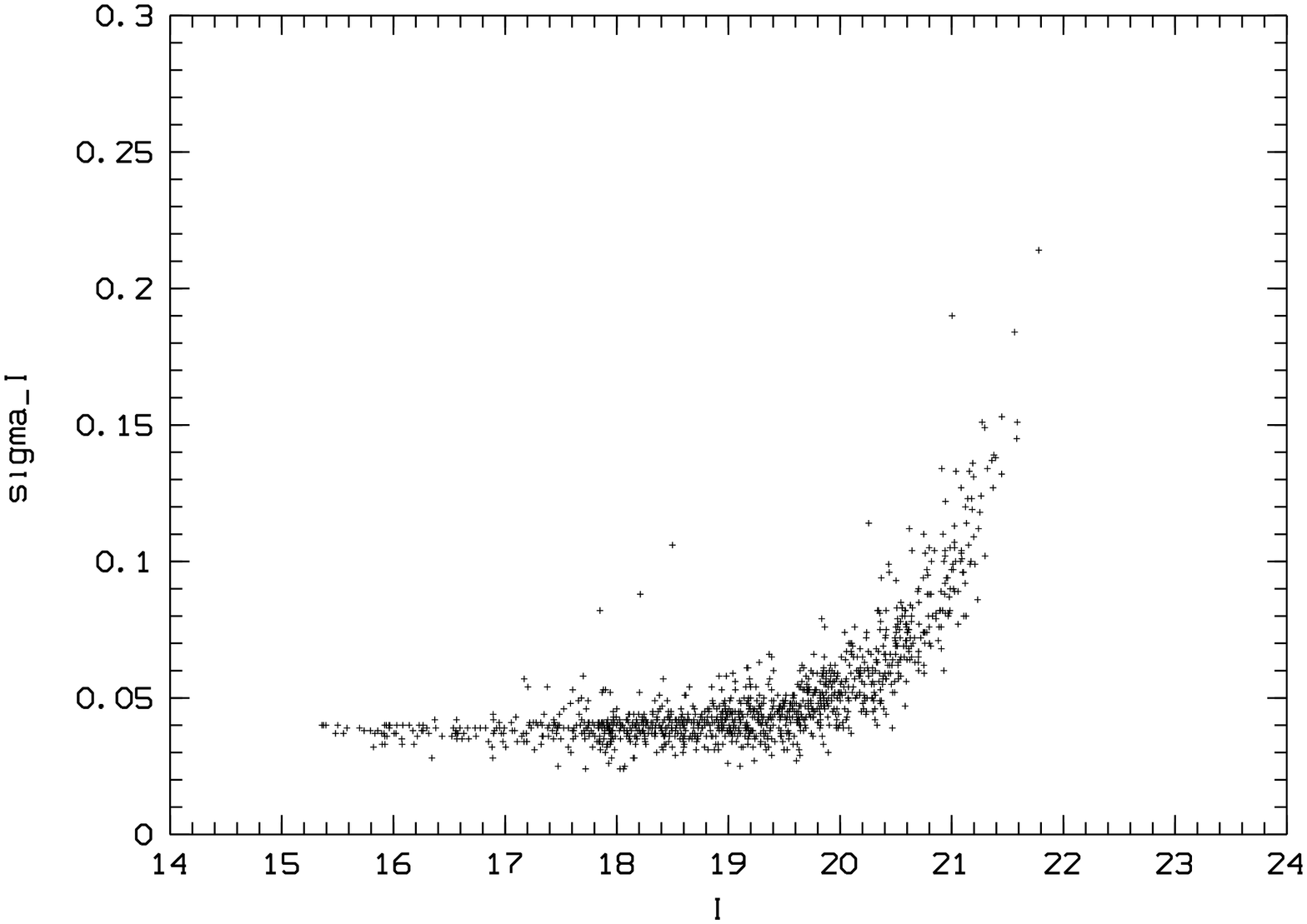}}
  \caption{Error distribution plots for the B, V and I photometry.}
\label{fig2}
\end{figure}

$I$ = 24.61 - 0.01 ($V-I$) + $v$

\noindent for 10 sec. and 15 sec. and 5 sec. in B, V, and I,
 respectively, at 1.1 airmasses. The cluster was observed
at an airmass of 1.25.
The errors in the zero point calibration are dominated
by the crowding in the transfer from aperture to convolved
magnitudes, of  about $\pm$0.03 mag in each colour.
The CCD shutter time uncertainty (0.3 sec) 
gives   an additional 3\% error.
The magnitude zero point uncertainty is around
 $\pm$0.05.  In Figs. \ref{fig2}a,b,c are shown the error
distribution plots for the B, V and I photometry. These figs.
show the standard errors, derived on the basis of
statistical poissonian noise produced by DAOPHOT.
For the atmospheric extinction we applied  the
 La Silla coefficients\footnote{http://www.ls.\-eso.org/\-lasilla/\-atm-ext/}.

\begin{table}
\caption[1]{Log of observations}
\begin{flushleft}
\begin{tabular}{llllllll}
\noalign{\smallskip}
\hline
\noalign{\smallskip}
Target &    Filter  & Exp. & Seeing \\
\noalign{\smallskip}
& & &  
${\rm (")}$ \\
\noalign{\smallskip}
\hline
\noalign{\vskip 0.2cm}
ESO 92-SC05 &       $V$     &        60     &        1.25 \\
      &             $V$      &      900    &        1.25 \\
     &              $B$     &       60      &       1.25 \\
     &              $B$      &      1800      &      1.25 \\      
           &         $I$    &        40    &        1.25 \\
           &         $I$    &        600         &    1.25 \\
\noalign{\smallskip}
\noalign{\smallskip} \hline \end{tabular}
\\
\end{flushleft} 
\end{table}

\begin{figure}
\resizebox{\hsize}{!}{\includegraphics[angle=-90]{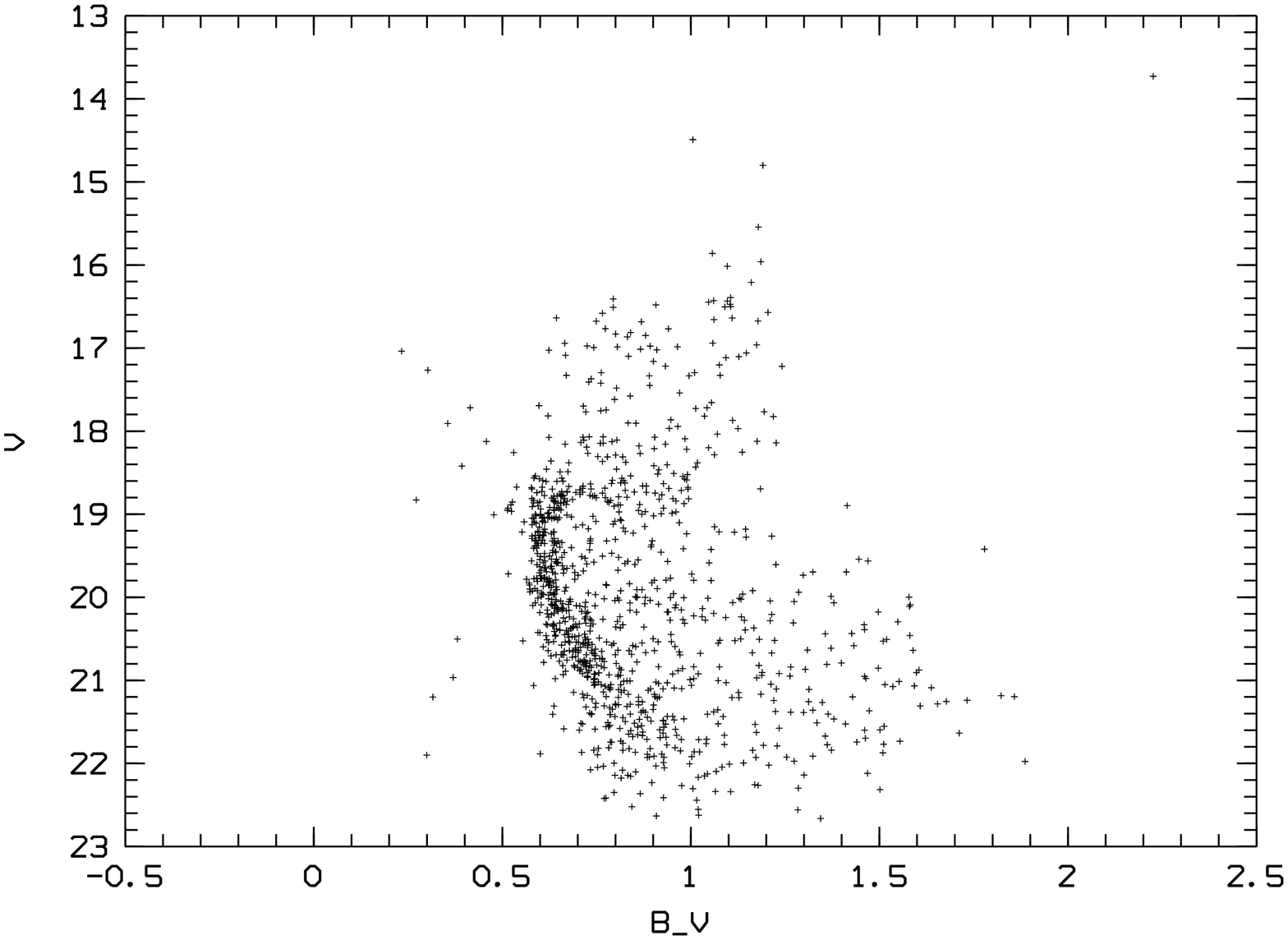}}
\hspace*{0.1cm}%
\resizebox{\hsize}{!}{\includegraphics[angle=-90]{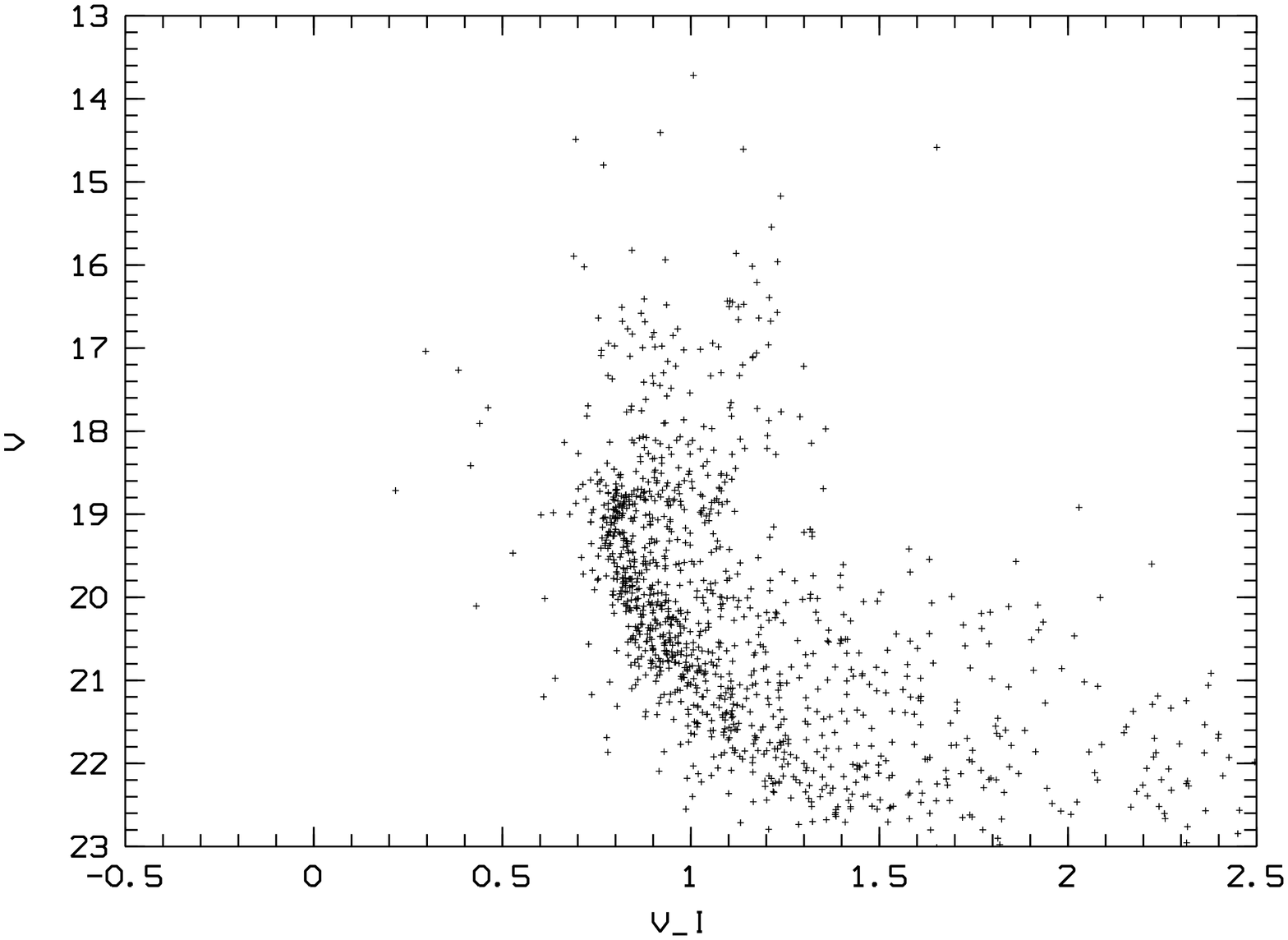}}
  \caption{
ESO92-SC05: (a) $V$ vs. $B-V$; (b) V vs. $V-I$ CMDs for an extraction of 
r$<$300 pixels (r$<$117'').
}
\label{fig3}
\end{figure}


\begin{figure*}[ht]
\resizebox{9.2cm}{!}{\includegraphics[angle=-90]{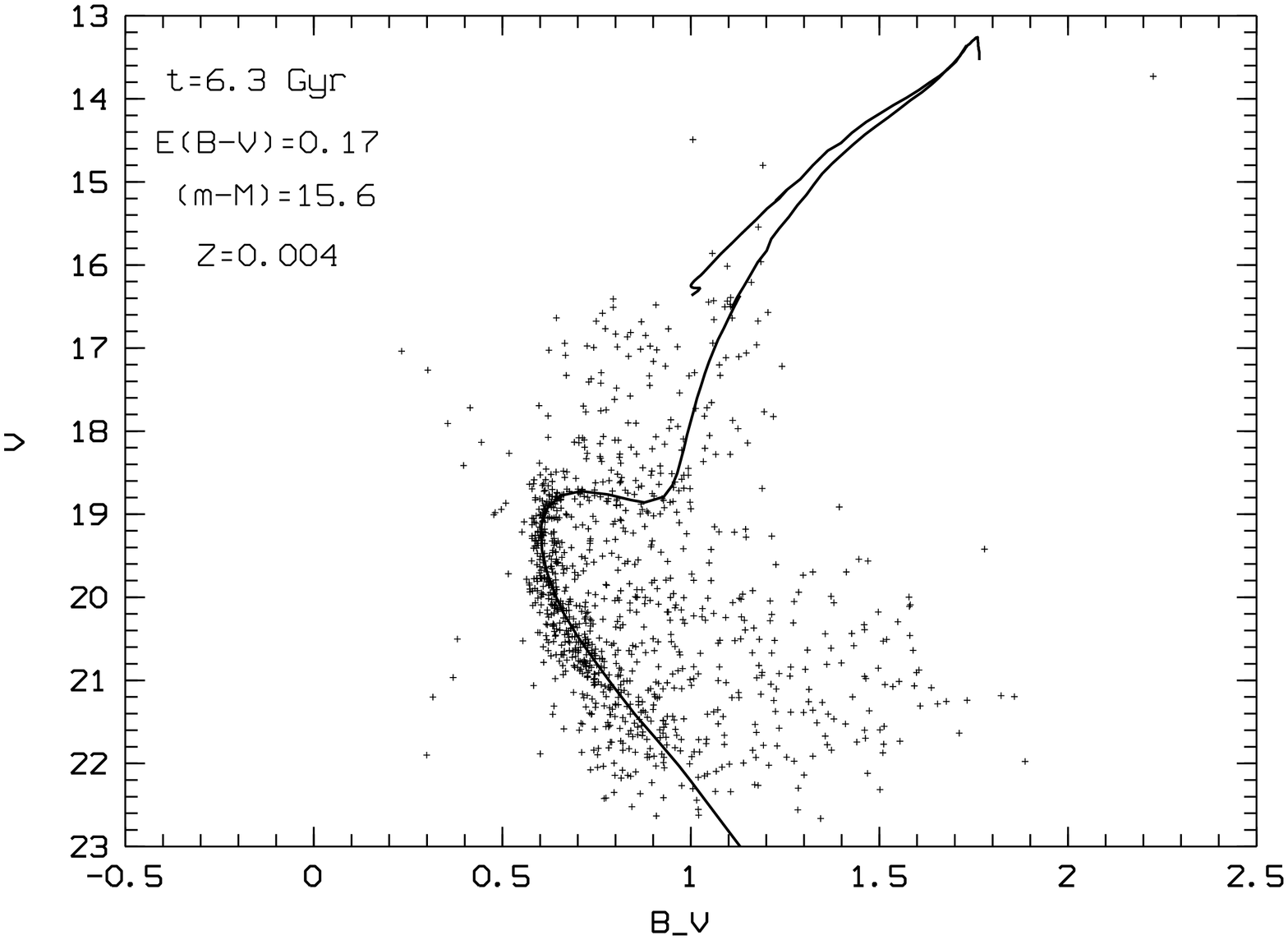}}%
\hspace*{0.1cm}%
\resizebox{9.2cm}{!}{\includegraphics[angle=-90]{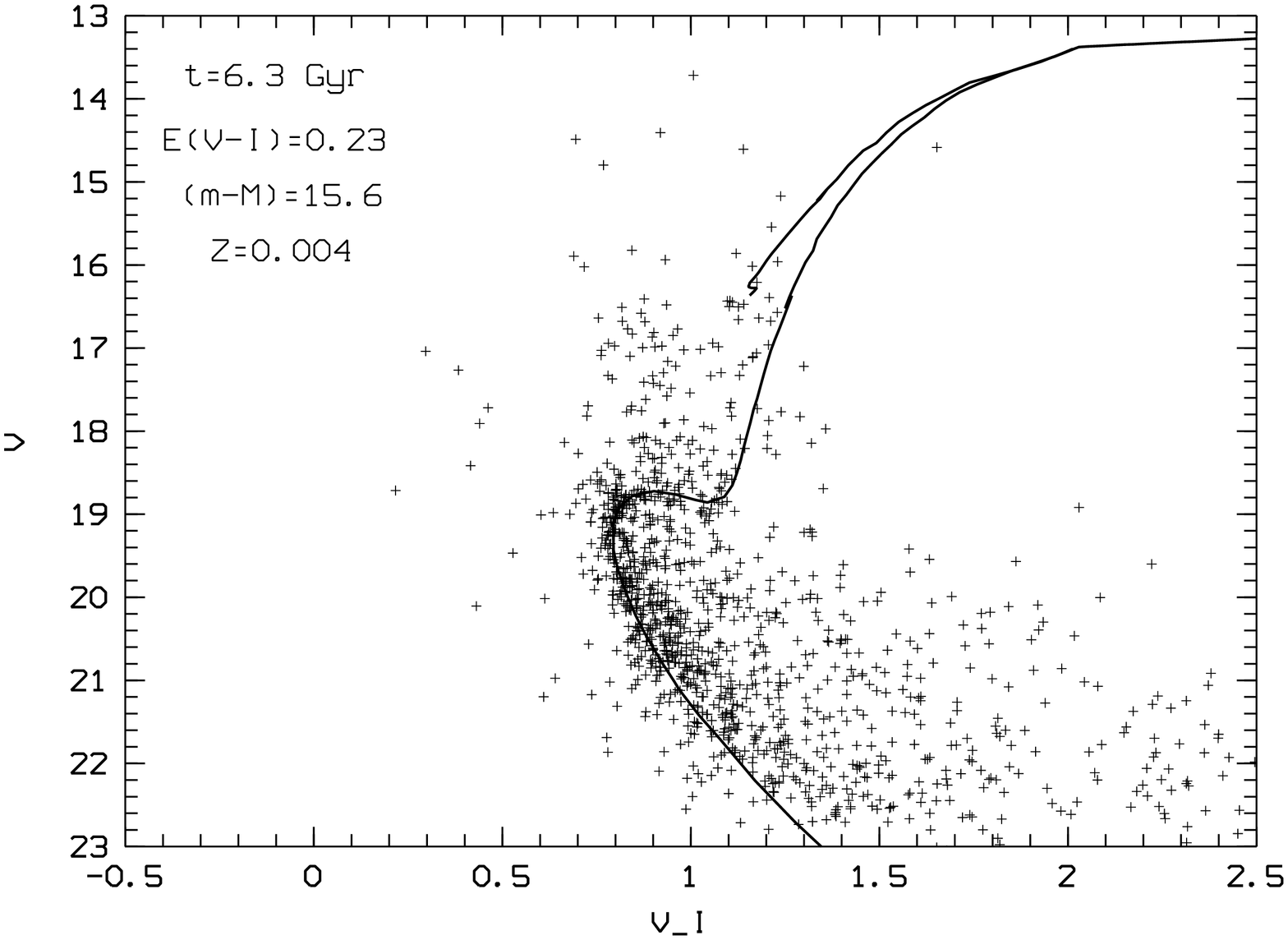}}
\caption{ESO 92-SC05:  (a) V vs. (B-V); (b) V vs. (V-I) for the
same extraction of Figs. 3a,b,  with Padova isochrones for
age=6.3 Gyr and metallicity Z=0.004 ([M/H]$\sim$-0.7) overplotted.}
\label{fig4}
\end{figure*}

\begin{figure}
\resizebox{\hsize}{!}{\includegraphics[angle=-90]{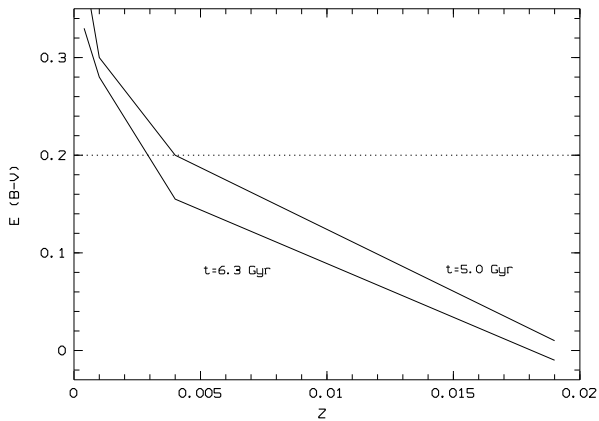}}
  \caption{
Reddening E(B-V) vs. metallicity Z for different ages.
The dotted line corresponds to the reddening of ESO92-SC05,
from Schlegel et al. (1998).
}
\label{fig5}
\end{figure}

\begin{figure}
\resizebox{\hsize}{!}{\includegraphics[angle=-90]{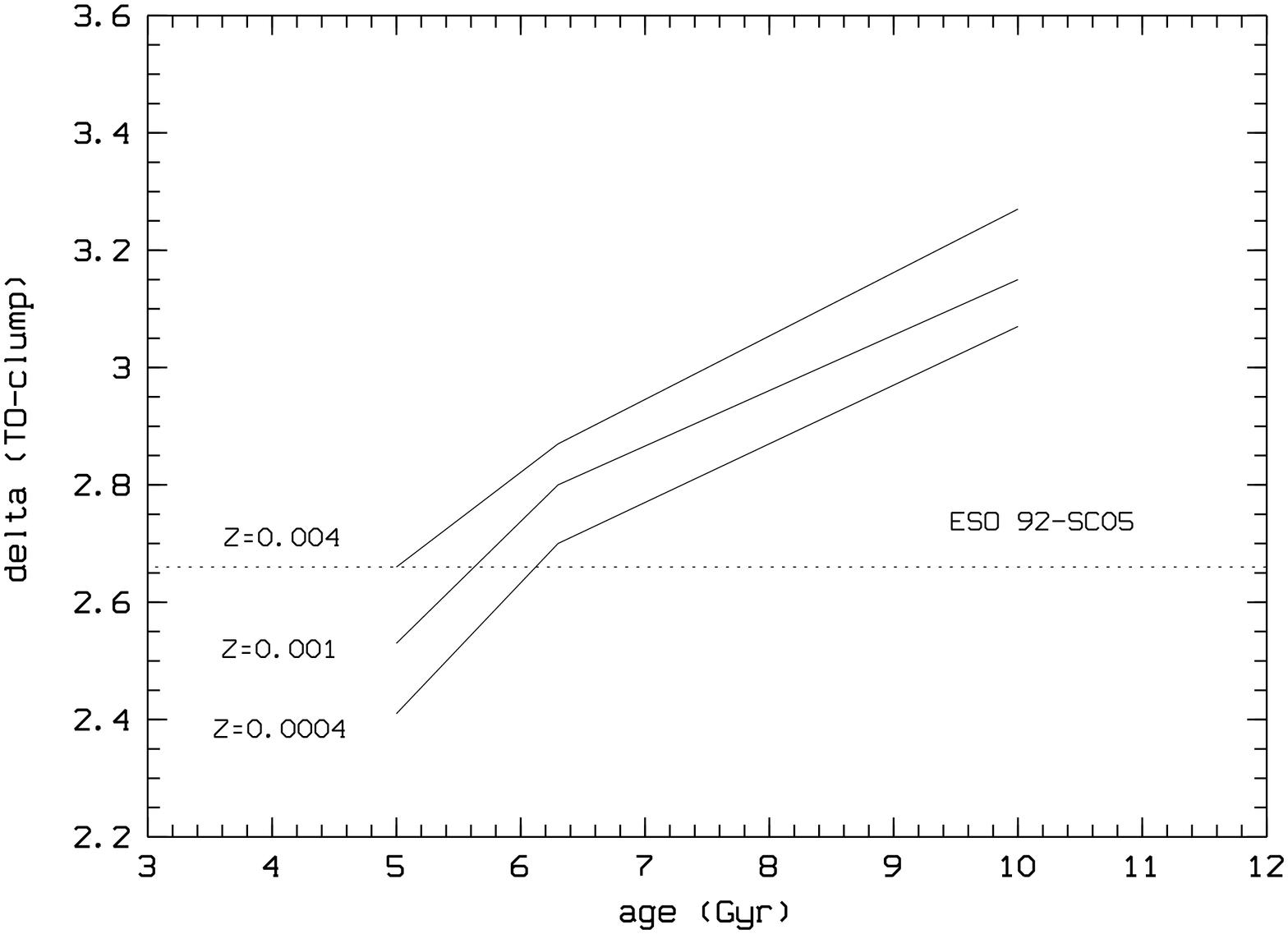}}
  \caption{
Magnitude difference between TO and clump $\Delta$V(TO-clump)
vs. age (Gyr) for different metallicities Z. The
dotted line corresponds to the $\Delta$V(TO-clump)
 value of ESO92-SC05.
}
\label{fig6}
\end{figure}

\section {Colour-Magnitude Diagrams}

In Figs. \ref{fig3}a,b the $V$ vs. $B-V$ and  $V$ vs. $V-I$
 CMDs of ESO 92-SC05  
are shown for an extraction of r $<$ 300 pixels (r$<$ 117'').

A well populated old turn-off (TO) is evident, as well as a clump
also suggesting an old age, and possibility of a few blue
stragglers. The Red Giant Branch (RGB) is steep
indicating a low metallicity. The shape of the TO and subgiant
branch (SGB) suggest an intermediate or low metallicity.

The clump is  located at $V$=16.45 and $B-V$=1.13 (or $V-I$=1.18). 
From the luminosity difference between the main sequence TO 
at $V$$\approx$19.1 and the clump, an age range between 5.0 and
6.3 Gyr is inferred by comparison with Padova isochrones
(Girardi et al. 2000).
The CMDs show sequences specially well defined at the level of the TO
and SGB, despite the presence of field contamination.
 The $V$ vs. $V-I$ CMD is deeper.

\section{Cluster parameters}

Padova isochrones from Bertelli et al. (1994),
and Girardi et al. (1996, 2000)\footnote{http://pleiadi.pd.astro.it}
were used  to derive cluster parameters,
by fitting them simultaneously on the B,V and V,I diagrams.
 We adopted
the relation E(V-I)=1.33E(B-V) (Dean et al. 1978).

 Figs. \ref{fig4}a,b show the $V$ vs. $B-V$ and V vs. $V-I$ CMDs
best fit with  Padova isochrones of
age=6.3 Gyr and metallicity Z=0.004 ([M/H]$\sim$-0.7).
Reddening values in the range E(B-V)=0.17 to 0.19 are
acceptable from the B,V and V,I diagrams.

This result appears to be robust, as a result of a
 series of fits with isochrones in the age
and metallicity ranges of 6.3$<$t(Gyr)$<$5, 
0.004$<$Z$<$0.0004, and reddening values   0.30$<$E(B-V)$<$0.10.

Note in Fig. \ref{fig4}a that the colour of the RGB clump appears redder than 
the isochrone, whereas the luminosity level does correspond to the isochrone.
We consider that 
the fit in luminosity is a more reliable reference,
 because the luminosity transformations
are much more accurate than the temperature-colour transformations.
The colour displacement of the RGB clump, of the order of
0.05-0.07 mag. in B-V, is compatible with the
transformation errors from theoretical to observed planes,
which is a problem inherent to the presently available
isochrone calculations, 
due to still missing opacities in the treatment  of cool stars atmospheres. 

From our comparisons, we concluded that,  firstly we cannot obtain a fit for
the same metallicity and a younger age because, even if the RGB colour would be
closer, the luminosity of the clump, in the case of 5 Gyr,
would be off by 0.2 mag in V.
Second, by  increasing the
metallicity, where a closest set to
our previous fit is obtained with solar metallicity (vs. our Z=0.004) 
and 5 Gyr, it reduces the discrepancy in V, but there is still a 0.15 mag.
discrepancy, while the colour of the RGB clump gets a good fit.
In this case the isochrones are however considerably redder 
and the colour excess
(reddening) drops to zero or to slightly negative values. Furthermore the two colours
(V-I and B-V) would give a different reddening, with E(V-I)=0.15 and the whole
V, I diagram is considerably off.
 Clearly only a lower metallicity is suitable.
.....

As concerns $\alpha$-enhancement derived for metal-poor open
clusters (e.g. Carraro et al. 2007), we found that there is 
no need trying to fit $\alpha$-enhanced
isochrones, given that Salaris et al. (1993) showed that the 
$\alpha$-enhancement effect on
the isochrones is equivalent to metallicity scaled isochrones. 
It is important to stress that the
 Salaris et al. results are valid for low metallicities, and fail
at high ones (near-solar and above), as found by
Salaris and Weiss (1998), and Salasnich et al. (2000).
For [Fe/H]$<$-0.5, in principle
solar-scaled isochrones can be safely used.

In Fig. \ref{fig5} the reddening E(B-V) 
is plotted against metallicity Z for different ages,
derived from the isochrone set used. The line 
corresponding to the reddening of ESO92-SC05 indicates a
 metallicity  Z=0.0035$\pm$0.001, or [M/H]$\approx$-0.7. 
Chen et al. (2003) presented a histogram of metallicities
for 118 open clusters for which ages and metallicities are available.
Typical metal-poor open clusters show [Fe/H]$\approx$-0.4 (e.g.
Gratton \& Contarini 1994), while ESO92-SC05 is more metal-poor.
Chen et al.'s Fig. 4 showing the metallicity histogram indicates
that open clusters with [Fe/H]$<$-0.5 are 
extremely rare.
Fig. \ref{fig6} shows 
the magnitude difference between TO and clump $\Delta$V(TO-clump)
vs. age (Gyr) for different metallicities Z. The
dotted line  with the $\Delta$V(TO-clump)=2.65
value of ESO92-SC05,
indicates that its age is 5.6$\pm$1.0 Gyr.

The adopted cluster parameters are given in Table 2.
\begin{table}
\caption[]{Parameters of ESO92-SC05.}
\label{tab2}
\renewcommand{\tabcolsep}{1.75mm}
\renewcommand{\arraystretch}{1.5}
\begin{tabular}{lccc}
\hline\hline
parameters & values \\
\hline
Age (Gyr) &  6.0$\pm$1.0 \\
$\lbrack$M/H$\rbrack$  & -0.7$\pm$0.2 \\
(m-M)$_{\rm V}$ & 15.71$\pm$0.1 \\
E(B-V) & 0.17$\pm$0.04 \\
A$_{\rm V}$ & 0.53$\pm$0.15\\
(m-M)$_{o}$ & 15.18$\pm$0.2\\
d$_{\odot}$  & 10.9$\pm$1.0 kpc \\
R$_{\rm GC}$ &  11.4$\pm$1.0 kpc \\
X$_{\rm GC}$ & 4.5$\pm$0.4kpc \\
Y$_{\rm GC}$ &-10.4$\pm$1.0kpc\\
Z$_{\rm GC}$ &1.4$\pm$0.2kpc\\
\hline
\end{tabular}
\end{table}

Until recently the adopted distance of the Sun to the Galactic center
was  R$_{\rm GC}$=8.0kpc,  as reviewed by Reid (1993). Various methods have 
given smaller values.  Eisenhauer et al. (2005) derived 
R$_{\rm GC}$=7.6$\pm$0.3 kpc,  Nishiyama (2006) 
found R$_{\rm GC}$=7.5$\pm$0.35 kpc, 
and Bica et al. (2006) R$_{\rm GC}$=7.2$\pm$0.3 kpc. 
 On the other hand, Groenewegen et al. (2008) obtained a 
longer distance (R$_{\rm GC}$=7.94 $\pm$0.37 kpc).
Recently, Nikiforov (2008)
has shown that the galactocentric distance from solving for
the stellar orbit around Sgr A is not as precise so far, due to
systematic errors. As a compromise value, we adopt
 R$_{\rm GC}$=7.5 kpc. The Galactocentric coordinates are shown 
in Table 2. For R$_{\rm GC}$=8 kpc the results are similar.

In a CMD of the total field of ESO 92-SC05, we counted 10-15 clump
stars, depending on contamination. M67 has $\approx$ 5 clump stars
and estimated to have 724 M$_{\odot}$ (Montgomery et al. 1993). 
 Fan et al. (1967) found 6 clump stars in M67 that were   kinematically
selected members.
They derived a total cluster mass of M=1270 M$_{\odot}$ 
(stars with masses larger
than 0.5 M$_{\odot}$) for R$<$66.6´(15.2 pc).
Within a comparable  spatial radius ESO92SC-05 has 8 clump stars leading to
a total mass M$\approx$1700 M$_{\odot}$.
From the ratio of clump stars, we estimate a mass
of 1800$\pm$400 M$_{\odot}$ for ESO 92-SC05.
Thus we are dealing with a massive open cluster in the outer disk.

\subsection{Age distribution of old open clusters}

\begin{figure}
\resizebox{\hsize}{!}{\includegraphics{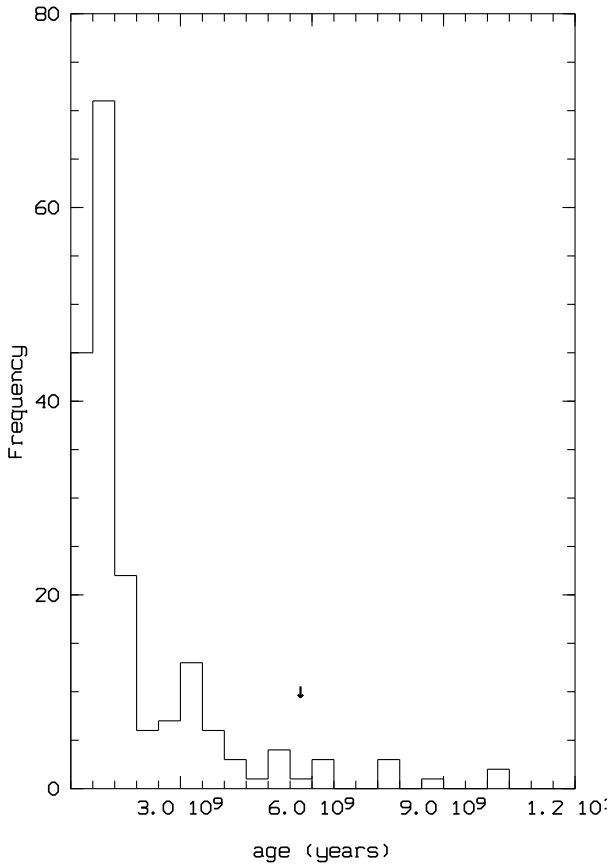}}
  \caption{
Age histogram for old open clusters based on
Dias et al. (2002) catalogue. The arrow indicates the age
bin of ESO92-SC05.
}
\label{fig7}
\end{figure}

Compilations of old open clusters are bringing 
progressively new objects (e.g. Frinchaboy \& Phelps 2002)
 and a clearer picture
of their spatial, age and metallicity distributions
(Friel 1995; Dutra \& Bica 2000; Ortolani et al. 2005b).
Currently, in Dias et al. (2002) catalogue
188 open clusters are reported to be older than 700 Myr.
Relative to Ortolani et al. (2005a) the sample increased
by about 80\%. In Fig. \ref{fig7} we show the histogram of the
188 old open clusters presently known, and now including ESO92-SC05.
It is clear from this Fig. that ESO92-SC05 is among a dozen
open clusters older than 6 Gyr so far known in the Galaxy.

 Open clusters in general
may be dissolved in a typical timescale of 600 Myr 
(Bergond et al. 2001), whereas for massive clusters
(m$\sim$10$^4$M$_{\odot}$), the timescale for disruption
is estimated to be 2 Gyr (Gieles et al. 2006).
This is compatible with the histogram, indicating
that the older clusters probably are a surviving
population of initially massive clusters.

\section {Conclusions}

We present BVI photometry of the star cluster ESO92-SC05.
A reddening E(B-V)=0.17 and a distance from the Sun 
d$_{\odot}$= 10.9 kpc are found.
 ESO92-SC05 is one of the few  distant, old open clusters studied
in detail in the fourth quadrant, along with Saurer 3,
ESO92-SC18, ESO93-SC08 and BH 144 (Dias et al. 2002).

 The old age of 6.0 Gyr for ESO92-SC05 includes it in
 the sample of rare Galactic open clusters
known to be older than 5 Gyr (Ortolani et al. 2005a,b;
see also the catalogue by Dias et al. 2002).

A low metallicity of [M/H]$\sim$-0.7 is found for ESO92-SC05.
Assuming
the metallicity gradient of -0.06dex kpc$^{-1}$ over the Galactocentric
distances of 7 to 16 kpc  derived by Friel et al. (2002),
and confirmed by Salaris et al. (2004), we get
an expected value [M/H]$\sim$-0.34 at the distance of 11kpc.
ESO 92-SC05 is thus of lower metallicity. It appears to be
among the most metal-poor open clusters in the Galaxy (Friel
et al. 2002; Chen et al. 2003).

The origin of this cluster is an interesting issue. It belongs
to outer disk. The distance from the plane of Z$_{\rm GC}$ 1.4 kpc
 is high. Is it a member of the outer disk that could be thick,
or  captured from an infalling dwarf galaxy (Yong et al. 2005;
Frinchaboy et al. 2004; Rocha-Pinto et al. 2006).
 ESO92-SC05 show characteristics similar to Berkeley 29 and
Saurer 1, which were considered to be possible accretion
products from the Galactic Anticenter
Stellar Strucure (GASS) by Frinchaboy et al. (2006).
In fact, ESO92-SC05's  location at (l=286.2$^{\circ}$,b=-7.5$^{\circ}$)  
is rather far from the proposed
direction (l=240$^{\circ}$,-8$^{\circ}$) and extent of the accreted
dwarf galaxy  Canis Major  (Martin et al. 2004). Moreover Canis Major has
been argued to be
part of the Galactic warp (Momany et al. 2006). On the other hand ESO92-SC05
projection in Carina  and its distance (Table 2) match
the Argo-Navis structure  (Rocha-Pinto et al. 2006). Finally, ESO92-SC05
matches the 4th quadrant extensions of GASS. 
It fits quite well not only spatially but also  the
age-metallicity relation, occupying the lower envelope of the distribution
(Frinchaboy et al. 2004).

\section*{Acknowledgments}

We are grateful to Leo Girardi for useful information on
the Padova isochrones.
We acknowledge partial financial support from
the Brazilian agencies  Fapesp and CNPq,
and the Italian Ministero dell'Universit\`a e della Ricerca
Scientifica e Tecnologica (MURST).

\end{document}